\def\la{\hbox{{\lower -2.5pt\hbox{$<$}}\hskip -8pt\raise
-2.5pt\hbox{$\sim$}}}
\def\ga{\hbox{{\lower -2.5pt\hbox{$>$}}\hskip -8pt\raise
-2.5pt\hbox{$\sim$}}}

\def\NPB{{\em Nucl. Phys.} B}

\def\PRL{\em Phys. Rev. Lett.}
\def\PRD{{\em Phys. Rev.} D}

\documentstyle[epsfig]{aipproc}

\begin{document}
\title{From the Galaxy to the Edge of the Universe: Plausible Sources of
UHECRs}

\author{Angela V. Olinto}
\address{Department of Astronomy \& Astrophysics, \\ \& Enrico Fermi
Institute,\\
The University of Chicago, Chicago, IL 60637}

\maketitle

\begin{abstract}
The lack of a high energy cutoff in the cosmic ray spectrum  
together with an apparently isotropic distribution of arrival
directions for the highest energy events have strongly constrained most
models proposed for the generation of these particles.  An overview of
the theoretical proposals are presented along with their  
most general signatures.  Future experimental tests of the different
proposals are discussed.
\end{abstract}

\section*{Introduction}

The surprising detection of cosmic rays with energies above $10^{20}$ eV 
has triggered considerable interest on  the  origin and nature of these
particles. In addition to the ultra-high energy events detected by  Fly's
Eye,  Haverah Park,  Yakutsk, and Volcano Ranch, the AGASA experiment has
recently reported many hundreds of events (728) accumulated with energies
above $10^{19}$ eV and 8 events above $10^{20}$ eV  \cite{haya00}. (For a
recent review of the observations, see, e.g., \cite{revdata}.) 

These observations are surprising because not only  the energy requirements
for astrophysical sources to accelerate particles to $> 10^{20}$ eV are
extraordinary, but the propagation of particles at these energies is prone
to large energy losses. Reactions of ultra-high energy proton, nuclei, or
photon primaries with cosmic background radiation in intergalactic space
suppress the observable flux at the highest energies significantly. 
In fact, cosmic ray protons of energies above  a few $10^{19}$ eV  reach
the $\Delta$ resonance threshold and  produce pions  off the
cosmic microwave background (CMB) limiting their source to lie not much 
further than about $50\,$Mpc away from Earth. This photopion production
gives rise to the well-known  Greisen-Zatsepin-Kuzmin (GZK) cutoff in the
spectrum of cosmic ray photons \cite{GZK66}. Nuclei  are
photodisintegrated on shorter distances due to the infrared background
\cite{PSB76SS99} while the radio background constrains photons to originate
from even closer systems \cite{bere70PB96}.

If these UHE particles are protons, they  likely originate in extragalactic
sources, since  at these high energies the Galactic magnetic field cannot
confine protons in the Galaxy. If protons are extragalactic, they traverse
large intergalactic distances so their spectrum should exhibit
the GZK cutoff. The shape of the GZK
cutoff depends on the source input spectrum and the distribution of
sources in space as well as in the intergalactic magnetic field. In
Figure 1, we contrast the observed flux by AGASA with the expected flux
for proton sources distributed homogeneously or distributed like 
galaxies with injection spectrum $J(E)
\propto E^{-\gamma}$ and $\gamma = 3$ \cite{Blasi00,BBO00}. We model the
distribution of UHECR sources by using the galaxy distribution
measured by the recent IRAS redshift survey know as PSCz 
\cite{saunders00a}.  As can be seen from the figure, even allowing for the
local overdensity the observations are consistently above the theoretical
expectation. In fact, when we normalize our simulations by requiring that
the number of events with $E
\ge 10^{19}$ eV equals the AGASA observations (728), we find that the
number of expected events for $E \ge 10^{20}$ eV is only $1.2 \pm 1.0$ for
the PSCz case, i.e., 6 $\sigma$ away from the observed 8 events.

 \begin{figure}[thb]
 \begin{center}
  \mbox{\epsfig{file=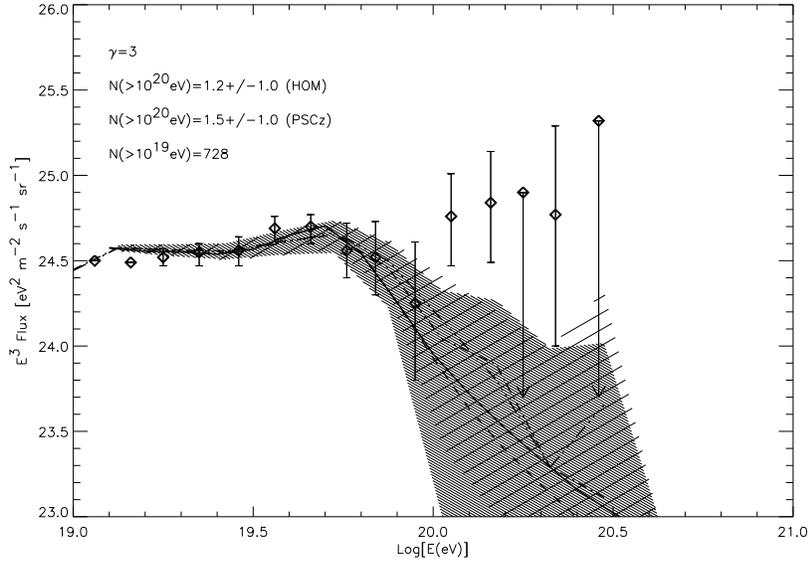,width=11cm}}
  \caption{\em {Simulated fluxes for the AGASA statistics of 728 events above 
$10^{19}$ eV, and $\gamma=3$, using a homogeneous source 
distribution ($\setminus$ hatches)
and the PSCz distribution (dense / hatches). The solid and dashed lines
are the results of the analytical calculations for the same two cases. 
The dash-dotted and dash-dot-dot-dotted lines trace 
the mean simulated fluxes for the homogeneous and the PSCz cases. 
(see [7]).}}
 \end{center}
\end{figure}

The gap between observed flux and model predictions narrows as the injection
spectrum of the ultra-high energy cosmic ray (UHECR) sources becomes much 
harder than $\gamma = 3$. The results for $\gamma=2.1$ are shown in Figure
2. For $\gamma = 2.1$, the number of expected events above $10^{20}$ eV
reaches $3.3 \pm 1.6$ for a homogeneous distribution while for the PSCz
catalog it is $3.7\pm 2.0$.  This trend can be seen also in Figure 3, where
analytic solutions for $\gamma =1.5, 2.1$ and 2.7 are shown.

\begin{figure}[thb]
 \begin{center}
  \mbox{\epsfig{file=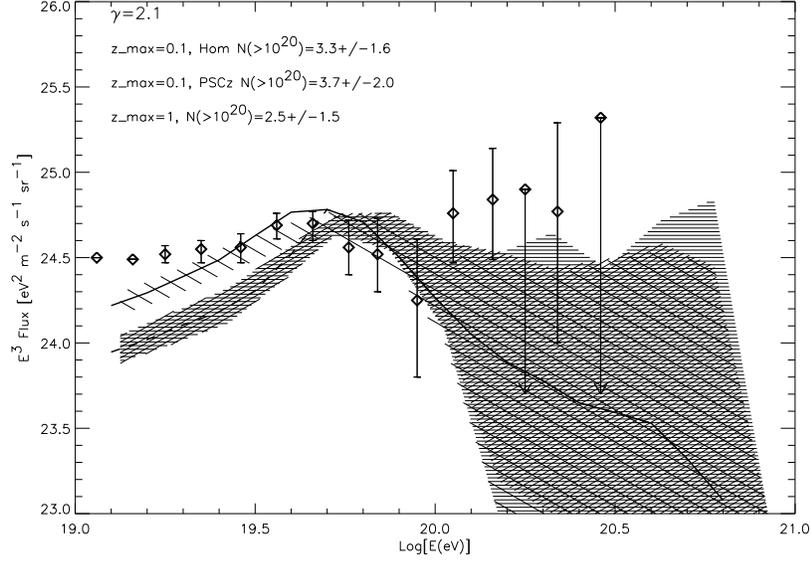,width=11cm}}
  \caption{\em {Simulated fluxes for the AGASA 
statistics of 728 events above 
$10^{19}$ eV, and $\gamma=2.1$, using a homogeneous source 
distribution with $z_{max}=0.1$ (/ hatches),
the PSCz distribution with $z_{max}=0.1$ (horizontal hatches), and
a homogeneous source 
distribution with $z_{max}=1$ ($\setminus$ hatches).
}}
 \end{center}
\end{figure}

In addition to the presence of events past the GZK cutoff, there has
been no clear counterparts identified in the  arrival direction of the
highest energy events. If these events are protons or photons,  these
observations should be astronomical, i.e., their arrival directions should
be the angular position of sources.  At these high energies the Galactic
and extragalactic magnetic fields do not affect proton orbits significantly
so that even protons should point back to their sources within a few degrees.
Protons at $10^{20}$ eV propagate mainly in straight lines as they traverse
the Galaxy since their gyroradii are $\sim $ 100 kpc in $ \mu$G  fields
which is typical in the Galactic disk. Extragalactic fields are expected to
be $\ll \mu$G \cite{KronVallee,BBO99}, and induce at most  
$\sim$ 1$^o$ deviation from the source. Even if
the Local Supercluster has relatively strong fields, the highest energy
events may deviate at most $\sim$ 10$^o$  
\cite{RKB98,SLB99}.  At present, no correlations between arrival directions
and plausible optical counterparts  such as sources in the Galactic
plane, the Local Group, or the Local Supercluster have been clearly
identified.  Ultra high energy cosmic ray data are consistent
with an isotropic distribution of sources in sharp contrast to the
anisotropic distribution of light within 50 Mpc from Earth.

The absence of a GZK cutoff and the isotropy of arrival directions are
two of the many challenges that models for the origin of UHECRs face.
This is an exciting open field, with many scenarios being proposed but 
no clear front runner.  Not only the origin of these 
particles  may be due to physics beyond the standard model of particle
physics, but their existence  can be used to constrain extensions of the
standard model such as violations of Lorentz invariance (see, e.g.,
\cite{ABGG00}). 

In the next section, we summarize the astrophysical Zevatron proposals while
models that involve new physics are discussed in the following section. To
conclude, future observational tests of UHECR models and their implications
are discussed.  (For recent reviews see
\cite{olinto00,BS00,Ber99,bland99}.)

 \begin{figure}[thb]
 \begin{center}
  \mbox{\epsfig{file=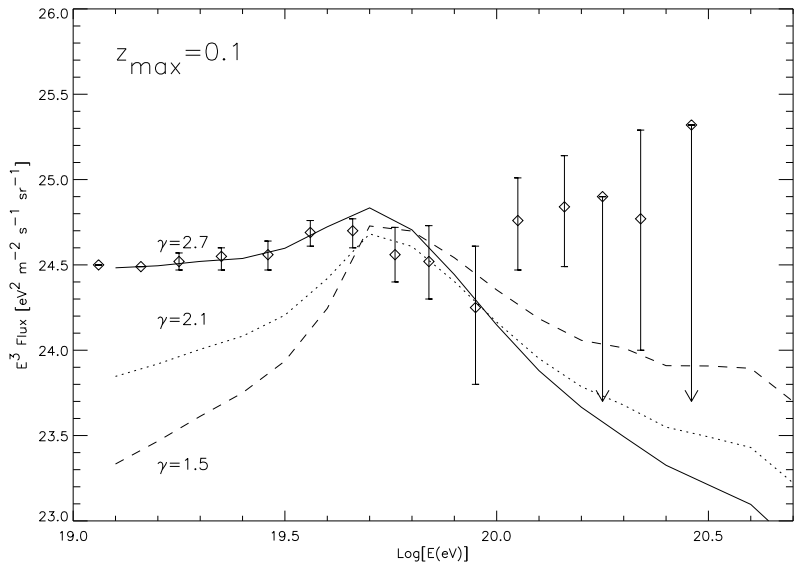,width=11cm}}
  \caption{\em {Propagated spectrum for source spectral index of $\gamma$ =
1.5, 2.1, 2.7}}
 \end{center}
\end{figure}

\section*{Astrophysical Zevatrons}

The puzzle presented by the observations of cosmic rays above $10^{20}$ eV
have generated a number of  proposals that we divide here as {\it
Astrophysical Zevatrons} and {\it New Physics} models.  Astrophysical
Zevatrons are also referred to as  bottom-up models and involve searching
for acceleration sites in known astrophysical objects that can reach ZeV
energies. New Physics proposals can be either hybrid or pure top-down
models. Hybrid models involve Zevatrons and extensions of the particle
physics standard model while top-down models involve the decay of very
high mass relics from the early universe and physics way beyond the standard
model. Here we discuss astrophysical Zevatrons while new physics models
are discussed in the next section. 

Cosmic rays can be accelerated in
astrophysical plasmas when large-scale macroscopic motions, such
as shocks, winds, and turbulent flows, are transferred to individual
particles. The maximum energy of accelerated  particles,
$E_{\rm max}$, can be estimated by requiring that the gyroradius of the
particle be contained in the acceleration region: $E_{\rm max} = Ze \, B
\, L$, where  $Ze$ is the charge of the particle, $B$ is the strength  and
$L$ the  coherence length of the magnetic field embedded in the plasma.
For $E_{\rm max} \ga 10^{20}$ eV and $Z \sim 1$, the only known
astrophysical sources with reasonable  $B L $ products   are neutron
stars ($B \sim 10^{13}$ G, $L \sim 10$ km),  active galactic nuclei (AGNs)
($B \sim 10^{4}$ G, $L \sim 10$ AU), radio lobes of AGNs ($B \sim
0.1\mu$G, $L \sim 10$ kpc), and clusters of galaxies ($B \sim \mu$G, $L
\sim 100$ kpc). In Figure 4, we highlight the  $B$ vs. $L$  for objects
that can reach  $E_{max} =  10^{20}$ eV with $Z=1$ (dashed line) and
$Z=26$  (solid line).   We discuss each of these candidates below.

\begin{figure}[thb]
 \begin{center}
  \mbox{\epsfig{file=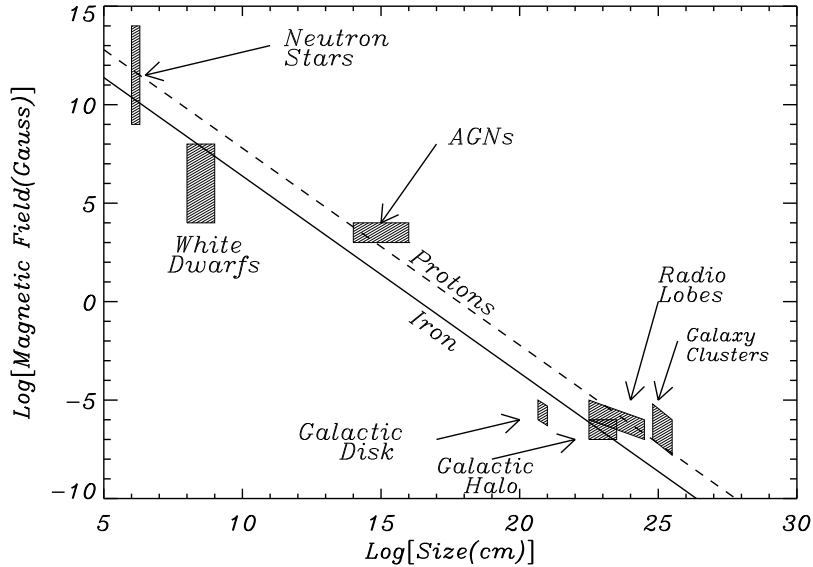,width=11cm}}
  \caption{\em {$B$ vs. $L$, for $E_{max} =  10^{20}$ eV, $Z=1$
(dashed line) and $Z=26$  (solid line) from [14].}}
 \end{center}
\end{figure}

{\it Clusters of Galaxies:}
Cluster shocks are
reasonable sites to consider for ultra-high energy cosmic ray (UHECR)
acceleration, since  particles with energy up to $E_{\rm max}$ can be
contained by cluster fields. However, efficient losses due to 
photopion production off the CMB during the propagation inside the
cluster limit UHECRs in cluster shocks  to reach at most
$\sim$ 10 EeV \cite{KRJ96KRB97}.

{\it AGN Radio Lobes:}
Next on the list of plausible Zevatrons
are extremely powerful radio galaxies  \cite{BS87B97}. 
Jets from the central black-hole of an active galaxy end at a termination
shock where the interaction of the jet with the intergalactic medium
forms radio lobes and  `hot spots'. Of special interest are the most
powerful AGNs where shocks can accelerate particles to energies well
above an EeV via the first-order Fermi mechanism. These sources may be
responsible for the flux of UHECRs up to the GZK cutoff \cite{RB93}.

A nearby specially powerful source may be able to reach energies  past the
cutoff.  However, extremely powerful AGNs with radio lobes
and hot spots are rare and far apart. The closest known object is M87 in
the Virgo cluster ($\sim$ 18 Mpc away)  and could be a main source of
UHECRs.  Although a single nearby source with especially hard spectra may 
fit the  spectrum  for a given strength and structure of the intergalactic
magnetic field \cite{BO99}, it is unlikely to match the observed arrival
direction distribution. If M87 is the primary source of UHECRs a
concentration of events in the direction of M87 or the Virgo cluster
should be seen in the arrival direction distribution. No such hot spot is
observed ({\it Hot spot} in Table 1). The next known nearby
source after M87 is NGC315 which is already too far at a distance of
$\sim $ 80 Mpc. Any unknown source between M87 and NGC315  would likely
contribute a second hot spot, not a isotropic distribution. The very
distant radio lobes will contribute a GZK cut spectrum ({\it dist RLs} in
Table 1).

The lack of a clear hot spot in the direction of M87 from the arrival
direction distribution has encouraged the idea that a strong Galactic
magnetic wind may exist that could help isotropize the arrival directions
of UHECRs.  A  Galactic wind with a  strongly magnetized azimuthal
component \cite{ABMS99} ($B_{GW}$ in Table 1) can
significantly alter the paths of UHECRs such that the observed arrival
directions of events above 10$^{20}$ eV would trace back to the North
Galactic pole which is close to the Virgo where M87 resides.
If our Galaxy has a such a wind is yet to be determined. The proposed wind
would focus most observed events into the
northern Galactic pole and render point source identification fruitless
\cite{BLS00}. Future observations of UHECRs from the Southern Hemisphere  
by the Southern Auger Site will provide precious data on previously
unobserved parts of the sky  and help   distinguish plausible proposals
for the effect of local magnetic fields on arrival directions.  Full sky
coverage is a key discriminator of such proposals.   

{\it  AGN - Central Regions:}
The powerful engines that give rise to the observed jets and radio
lobes are located in the central regions of active galaxies and are
powered by the accretion of matter onto supermassive black holes. It
is reasonable to consider the central engines themselves as the likely
accelerators \cite{T86,reviews}. In principle, the nuclei of  generic
active galaxies (not only the ones with radio lobes) can accelerate
particles via a unipolar inductor not unlike the one operating in
pulsars. In the case of AGNs,   the magnetic field  ($B_{source}$ in Table
1) may be provided by the infalling matter and the spinning black hole
horizon provides the imperfect conductor for the unipolar induction. 

The problem with AGNs as UHECR sources is two-fold: first, UHE particles
face  debilitating losses in the acceleration region due to the intense
radiation field present in AGNs,  and second, the
spatial distribution of objects should give rise to a  GZK cutoff of the
observed spectrum. In the central  regions of AGNs, loss processes are
expected to downgrade particle energies well below the maximum
achievable energy. This limitation has led to the proposal that  quasar
remnants, supermassive black holes in centers of inactive galaxies,  are
more effective UHECR accelerators \cite{BG99}. In this case, losses at the
source are not as significant but the propagation from source to us should
still lead to a clear GZK cutoff since sources would be associated with the
large scale structure of the galaxy distribution ({\it LSS} in Table 1).
From Figure 1--3, these models can only succeed if  the source spectrum
is fairly hard ($\gamma \la 2$) \cite{BBO00}.

\begin{table}[b!]
\caption{ZEVATRONS}
\label{table1}
\begin{tabular}{lddddd}
   Zevatron& 
   \multicolumn{1}{c}{Composition}&
   \multicolumn{1}{c}{Source $\gamma$} &
   \multicolumn{1}{c}{Sky Distrib.} &
   \multicolumn{1}{c}{$B$ Needs} &
   \multicolumn{1}{c}{Best Tests}\\
\tableline
Radio Lobes & Proton& 2--3 &  M87 +dist RLs  & $B_{GW}$ & Hot spot \&
$\gamma$\\ 
AGN Center & Proton & 2--3 & LSS  & $B_{source}$ & GZK feature \\
YNSWs & Iron  &1 & Gal. Disk  &  $B_{gal}$ &  Iron \& Disk \\ 
GRBs & Proton & 2--3 &Hot spot or  & large $B_{IGM}$
& Hot spot \& Flux\\
\end{tabular}
\end{table}

 {\it  Neutron Stars:}
Another astrophysical system capable of accelerating UHECRs is a neutron
star. In addition to having the ability to confine $10^{20} eV$ protons
(Figure 4), the rotation energy of young neutron stars is more than
sufficient to match the observed UHECR fluxes \cite{VMO97}.
However, acceleration processes inside the neutron star light cylinder are
bound to fail much like the AGN central region case:  ambient magnetic
and radiation fields induce significant losses. However, the plasma that
expands beyond the light cylinder is free from the main loss processes
and may be accelerated to ultra high energies.

One possible source of UHECR past the GZK cutoff is the early evolution
of neutron stars. In particular, newly formed, rapidly rotating
neutron stars may accelerate iron nuclei  to UHEs  through relativistic
MHD winds beyond  their light cylinders \cite{BEO99}. This mechanism
naturally leads to vary hard injection spectra ($\gamma \simeq 1$) (see 
Table 1). As seen in Figure 3,   $\gamma \sim 1$ improves the agreement
between predicted flux and observations for energies above $10^{20}$ eV. In
this case, UHECRs originate mostly in the Galaxy and the arrival directions
require that the primaries be  heavier nuclei. Depending on the structure
of   Galactic magnetic fields, the trajectories of iron nuclei from Galactic
neutron stars may be consistent with the observed arrival directions of the
highest energy events
\cite{ZPPR98}. Moreover,  if  cosmic rays  of a few times $10^{18}$ eV are
protons of Galactic origin, the isotropic distribution observed at these
energies is indicative of the diffusive effect of the Galactic magnetic
fields on iron at $\sim 10^{20}$ eV. This proposal should be constrained
once the primary composition is clearly determined (see {\it Iron} in
Table 1).

It has also been suggested that young extragalactic highly
magnetized neutron stars (magnetars) may be sources of UHE protons which
are accelerated by reconnection events \cite{GL00}. These would be prone
to a GZK cut spectrum and would need a very hard injection spectrum to
become viable explanations. 

{\it  Gamma-Ray Bursts:}
Transient high energy phenomena such as gamma-ray
bursts (GRBs) may also be a source of ultra-high energies protons
\cite{WV95}. In addition to both phenomena having unknown origins, GRBs
and UHECRs have other similarities that may argue for a common source.
Like UHECRs, GRBs are distributed isotropically in the sky,  and
the average rate of $\gamma$-ray energy emitted by GRBs is comparable
to the energy generation rate of UHECRs of energy $>10^{19}$ eV in a
redshift independent cosmological distribution of sources, both have  $ 
\approx 10^{44}{\rm erg\ /Mpc}^{3}/{\rm yr}.$ 

However, recent GRB counterpart
identifications argue for a strong cosmological evolution for GRBs. 
 The redshift dependence of GRB distribution is such that the flux of
UHECR associated with nearby GRBs would be too small to fit the
UHECR observations \cite{Ste99}. In addition, the distribution of UHECR
arrival directions and arrival times argues against the GRB--UHECR common
origin. Events past the GZK cutoff require that only GRBs from $\la 50$
Mpc contribute. Since less than about {\it one} burst is expected to have
occurred within this region over a period of 100 yr, the unique source
would appear as a concentration of UHECR events in a small part of
the sky (a {\it Hot spot} in Table 1).   In addition, the signal would be
very narrow in energy  $\Delta E/E\sim1$. Again, a strong intergalactic
magnetic field can ease the some of these difficulties giving a very
large dispersion in the arrival time and direction of protons  produced in
a single burst  ({\it large} $B_{IGM}$ in Table 1) \cite{WV95}.  Finally,
if the observed small scale clustering of arrival directions is confirmed
by future experiments with clusters having some lower energy events
clearly precede higher energy ones, bursts would be invalidated
\cite{SLO97}. 

\section*{New Physics Models}

The UHECR puzzle has inspired a number of different models that involve
physics beyond the standard model of particle physics. New Physics
proposals can be top-down models or a hybrid of astrophysical Zevatrons
with new particles. Top-down models involve the decay of very high mass
relics that could have been formed in the early universe.

The most economical among hybrid proposals involves a familiar
extension of the standard model, namely, neutrino masses.  If
some flavor of neutrinos have mass (e.g., $\sim 0.1 eV$), the relic
neutrino background  is a target for extremely high energy neutrinos to
interact and generate other particles by forming a Z-boson that
subsequently decays \cite{We97FMS97} (see $\nu$ {\it Z burst} in Table
2). If the universe has very luminous sources (Zevatrons) of extremely
high energy neutrinos ($\gg 10^{21}$ eV), these neutrinos would traverse
very large distances before annihilating with neutrinos in the smooth
cosmic neutrino background.  The UHE
neutrino Zevatrons can be much further than the GZK limited volume,
since neutrinos do not suffer the GZK losses. But if the interaction
occurs throughout a large volume, the GZK feature should also be
observed.  For plausible neutrino masses $\sim 0.1 eV$, the neutrino
background is very unclustered, so the arrival direction for events
should be isotropic and small scale clustering may be a strong challenge
for this proposal.  The weakest link in this proposal is the nature of a
Zevatron powerful enough to accelerate protons above tens of ZeVs that
can produce ZeV neutrinos as secondaries. This Zevatron is quite
spectacular, requiring an energy generation in excess of presently known
highest energy sources (referred to as {\it
$\nu$Zevatron} in Table 2).

Another suggestion is that the UHECR primary is a new hadronic
particle that is also accelerated in Zevatrons. The mass of a hypothetical
hadronic primary can be limited by the shower development of the Fly's
Eye highest energy event to be below $\la 50$ GeV \cite{AFK98}.  As in
the Z-burst proposal, a neutral particle  is usually harder to accelerate
and are usually created as secondaries of even higher energy charged
primaries. But once formed these can traverse large distances without
being affected by cosmic magnetic fields. Thus, a signature for future
experiments  of hybrid models that invoke new particles as primaries is a
clear correlation between the position of powerful Zevatrons in the sky
such as distant compact radio quasars and the arrival direction of  UHE
events \cite{FB98}. Preliminary evidence for such a correlation has been
recently reported \cite{VBJRRM}.       

Another exotic primary that can be accelerated to ultra high energies by
astrophysical systems is the vorton. Vortons are small loops of
superconducting cosmic string stabilized by the angular momentum of
charge carriers \cite{DS89}. Vortons can be a component of the dark
matter in galactic halos and be accelerated by astrophysical magnetic
fields \cite{BP97}. Vortons as primaries can be constrained by the
observed shower development profile.

It is possible that none of the astrophysical scenarios or the hybrid new
physics models are able to explain present and  future UHECR
data. In that case,  the alternative is to consider top-down models.
Top-down models involve  the decay of monopole-antimonoploe pairs
\cite{H83HS83}, ordinary and superconducting cosmic strings,  cosmic
necklaces, vortons, and superheavy long-lived relic particles.  The idea
behind these models is that relics of the very early universe,
topological defects (TDs) or superheavy relic (SHR) particles,  produced 
after or at the end of inflation, can decay today and generate UHECRs. 
Defects, such as cosmic strings, domain walls, and magnetic monopoles, 
can be generated through the Kibble mechanism  as symmetries are broken
with the expansion and cooling of the universe.  Topologically stable
defects can survive to the  present and decompose into their constituent
fields  as they collapse,  annihilate, or reach critical current in the
case of superconducting cosmic strings. The decay products, superheavy
gauge and higgs bosons, decay into jets of hadrons, mostly pions.  Pions
in the jets subsequently decay into $\gamma$-rays, electrons, and
neutrinos. Only a few percent of the hadrons are expected to be nucleons.
Typical features of these scenarios are a predominant release of
$\gamma$-rays and neutrinos and a QCD fragmentation spectrum which is 
considerably harder than the case of Zevatron shock acceleration.  

ZeV energies are not a challenge for top-down models since symmetry
breaking scales at the end of inflation typically are $\gg 10^{21}$
eV (typical X-particle masses vary between 
$\sim 10^{22-25}$ eV).  Fitting the observed flux
of UHECRs is the real challenge since the typical distances between TDs
is  the  Horizon scale about several Gpc. The low flux hurts proposals
based on ordinary  and superconducting cosmic strings which are
distributed throughout  space ({\it Distant TD} in Table 2). Monopoles
usually suffer the opposite problem, they would in general be too
numerous. Inflation succeeds in diluting the number density of
monopoles  and makes them too rare for UHECR production. To reach
the observed UHECR flux, monopole models usually involve some degree of
fine tuning. If enough monopoles and antimonopoles survive from the early
universe,  they may form a bound state, named monopolonium, that  can 
decay  generating UHECRs. The lifetime of monopolonia may be too short
for this scenario to succeed unless they are connected by strings
\cite{PO99}.

\begin{figure}[thb]
 \begin{center}
  \mbox{\epsfig{file=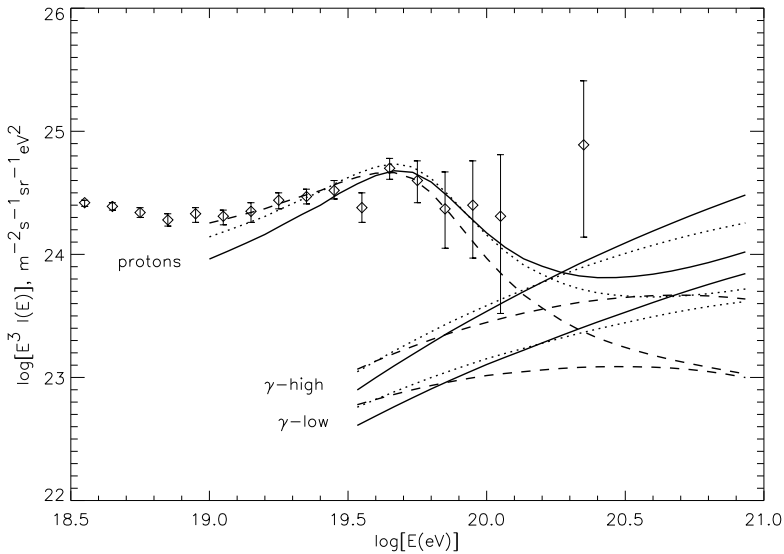,width=11cm}}
  \caption{\em {Proton and $\gamma$-ray fluxes from necklaces for 
$m_X=  10^{14}$ GeV (dashed lines),  $10^{15}$ GeV (dotted 
lines), and $10^{16}$ GeV (solid lines) normalized to
the  observed data.
$\gamma$-high  and  $\gamma$-low  correspond to two extreme cases 
of $\gamma$-ray absorption (see, [35]).
}}
\end{center}
\end{figure}

Once two symmetry breaking scales are invoked, a combination of
horizon scales gives room to reasonable number densities. This can be
arranged for cosmic strings that end in monopoles making a monopole
string network or even more clearly for cosmic necklaces \cite{BV97}.
Cosmic necklaces are hybrid defects where each monopole is connected to
two strings resembling beads on a cosmic string necklace. Necklace
networks may evolve to configurations that can fit the UHECR flux which
is ultimately generated by the annihilation of monopoles with
antimonopoles trapped in the string \cite{BV97,BBV98}. In these
scenarios, protons dominate the flux in the lower energy side of the GZK
cutoff  while photons tend to dominate at higher energies depending on
the radio background (see Figure 5 and {\it Distant TD} in Table 2). If 
future data can settle the composition of UHECRs from 0.01 to 1 ZeV, these
models can be well constrained. In addition to fitting the UHECR flux,
topological defect models are constrained by limits on the flux of  high
energy photons, from 10 MeV to 100 GeV, observed by EGRET.

\begin{figure}[thb]
 \begin{center}
  \mbox{\epsfig{file=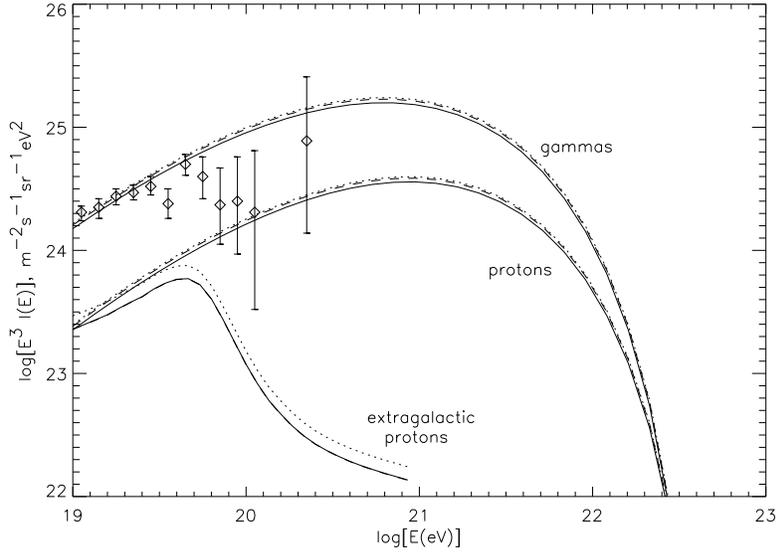,width=11cm}}
  \caption{\em {SHRs  or monopolia decay fluxes
(for $m_X= 10^{14} ~GeV$):
nucleons from the halo ({\it protons}), $\gamma$-rays
from the halo ({\it gammas}) and extragalactic protons. Solid, dotted
and dashed curves correspond to different  model parameters
(see [35]).
}}
\end{center}
\end{figure}

Another interesting possibility is the recent proposal that UHECRs are
produced by the decay of unstable superheavy relics that live much longer
than the age of the universe \cite{BKV97}.  SHRs may be produced at
the end of inflation by non-thermal effects such as a varying
gravitational field, parametric resonances during preheating,  instant
preheating, or the decay of topological defects. 
These models need to invoke special symmetries to insure unusually long
lifetimes for SHRs and that a sufficiently small percentage decays today
producing UHECRs \cite{BKV97,CKR99KT99}.  As in the topological defects
case, the decay of these relics also generates jets of hadrons.  These
particles behave like cold dark matter and could constitute a fair
fraction of the halo of our Galaxy. Therefore, their halo decay products
would not be limited by the GZK cutoff allowing for a large flux at UHEs
(see Figure 6 and {\it SHRs} in Table 2). Similar signatures can occur if
topological defects are microscopic, such  as monopolonia and vortons, and
decay in the Halo of our Galaxy ({\it Local TD} in Table 2). In both cases
({\it SHRs} and {\it Local TD}) the composition of the primary would be a
good discriminant since the decay products are usually dominated by photons.

\begin{table}[b!]
\caption{NEW PHYSICS}
\label{table1}
\begin{tabular}{lddddd}
   Source& 
   \multicolumn{1}{c}{Composition}&
   \multicolumn{1}{c}{Source $\gamma$}&
   \multicolumn{1}{c}{Sky Distrib.}&
   \multicolumn{1}{c}{Th. Challenge}&
   \multicolumn{1}{c}{Best Tests}\\
\tableline
   $\nu$ Z burst & photons   & $\nu$Zevatron &Isotropic  &  $\nu$Zevatron
 & photon, Isotropy \\   
   Distant TD & phot/GZK p & QCD frag & Isotropic & flux  &photon, GZK p\\
   Local TD & photons & QCD frag &Gal Halo  & origin &photon, Halo \\
   SHRs & photons & QCD frag &Gal Halo  & lifetime &photon, Halo \\ 
\end{tabular}
\end{table}

Future experiments should be
able to probe these hypotheses.  For instance, in the case of SHR
and monopolonium decays, the arrival
direction distribution should be close to isotropic but show an
asymmetry due to the position of the Earth in the Galactic
Halo \cite{BBV98} and the clustering due to small scale dark matter
inhomogeneities \cite{Blasi00,BlSe00}. Studying plausible halo models for
their expected asymmetry and inhomogeneitis will help constrain halo
distributions especially when larger data sets are available in the future.
High energy gamma ray experiments such as GLAST will also help constrain 
SHR models via the electromagnetic decay products \cite{B99}.

\section*{Conclusion}

Next generation experiments  such as the High Resolution Fly's Eye which
recently started operating, the Pierre Auger Project which is now under
construction,  the proposed  Telescope Array, and
the EUSO and OWL  satellites  will 
significantly improve the data at the extremely-high end of the cosmic
ray spectrum \cite{revdata}. With these observatories a clear
determination of the spectrum and spatial distribution of UHECR
sources is within reach. 

\begin{figure}[thb]
 \begin{center}
  \mbox{\epsfig{file=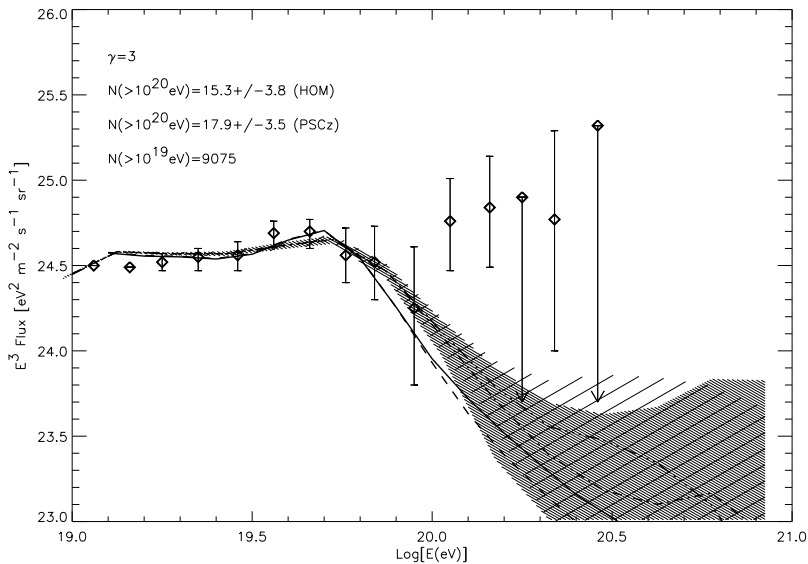,width=11cm}}
  \caption{\em {Simulated fluxes for the Auger projected 
statistics of 9075 events above 
$10^{19}$ eV, and $\gamma=3$, using a homogeneous source 
distribution ($\setminus$ hatches)
and the PSCz distribution (/ hatches). The solid and dashed
lines are the results of the analytical calculations for the same two
cases.  The dash-dotted and dash-dot-dot-dotted lines trace 
the mean simulated fluxes for the homogeneous and the PSCz cases. 
(see [7]).}}
 \end{center}
\end{figure}

The lack of a GZK cutoff should become clear with HiRes and Auger
and most extragalactic Zevatrons may be ruled out. 
 The observed spectrum will distinguish Zevatrons from
new physics models by testing the hardness of the spectrum and the
effect of propagation.  Figure 7 shows how clearly Auger will test the
spectrum in spite of clustering properties. The cosmography of sources
should also become clear and able to
 discriminate  between plausible populations for UHECR sources. 
The  correlation of arrival directions  for events with energies above
$10^{20}$ eV  with some known structure such as the Galaxy, the
Galactic halo, the Local Group or the Local Supercluster would be key
in differentiating between different models. For instance, a
correlation with the Galactic center and  disk should become apparent
at extremely high energies for the case of young neutron star winds,
while a correlation with the large scale galaxy distribution should
become clear for the case of quasar remnants. If SHRs or monopolonia are
responsible for UHECR production, the arrival directions should correlate
with the dark matter distribution and show the halo asymmetry. For these
signatures to be tested, full sky coverage is essential. Finally,  an
excellent discriminator would be an unambiguous composition
determination  of the primaries. In general, Galactic disk models  invoke
iron nuclei to be consistent with the isotropic distribution, 
extragalactic Zevatrons tend to favor proton primaries, while photon
primaries are more common for early universe relics. The hybrid detector
of the Auger Project should help determine the composition by measuring
the depth of shower maximum and the muon content of the
same shower.  The prospect of testing extremely high energy physics as
well as solving the UHECR mystery awaits improved observations that
should be coming in the next decade with experiments under construction
such as Auger \cite{Auger} or in the planning stages such as the Telescope
Array \cite{TA}, EUSO \cite{EUSO}, and OWL
\cite{OWL}.

\section*{Acknowledgment}
 
Many thanks to the organizers of this excellent workshop. This work was
supported by NSF through grant AST-0071235  and DOE grant DE-FG0291
ER40606.

\end{document}